\documentclass[manuscript,screen,authorversion]{acmart}
\usepackage[nameinlink,noabbrev]{cleveref}
\newgeometry{textwidth=6.05in}

\AtBeginDocument{%
  \providecommand\BibTeX{{%
    \normalfont B\kern-0.5em{\scshape i\kern-0.25em b}\kern-0.8em\TeX}}}

\setcopyright{rightsretained}
\copyrightyear{2021}
\acmYear{2021}

\acmConference[ACM RecSys SimuRec Workshop]{RecSys '21 SimuRec Workshop}{2021}{Amsterdam, Netherlands}
\acmBooktitle{SimuRec: Workshop on Synthetic Data and Simulation Methods for Recommender Systems Research
  2021 (at ACM RecSys '21), Amsterdam, Netherlands}

\begin{document}

\title{Recommendation System Simulations: A Discussion of Two Key Challenges}

\author{Allison J.B. Chaney}
\email{ajb.chaney@duke.edu}
\affiliation{%
  \institution{Duke University}
  \city{Durham}
  \state{North Carolina}
  \country{USA}
}

\begin{abstract}

As recommendation systems become increasingly standard for online platforms, simulations provide an avenue for understanding the impacts of these systems on individuals and society.
When constructing a recommendation system simulation, there are two key challenges: first, defining a model for users selecting or engaging with recommended items and second, defining a mechanism for users encountering items that are not recommended to the user directly by the platform, such as by a friend sharing specific content.
This paper will delve into both of these challenges, reviewing simulation assumptions from existing research and proposing alternative assumptions.  We also include a broader discussion of the limitations of simulations and outline of open questions in this area.
\end{abstract}


\maketitle

\section{Introduction}

With over 4.6 billion active internet user worldwide \citep{digitalpop}, online platforms have the potential to reach consumers at unprecedented rates.  Major online platforms typically provide personalized recommendations for their users---these system tailor search results, suggest products to purchase, customize front pages for news and opinions sites, curate social media posts, recommended movies, and generate music playlists.  Basic recommendation systems are relatively straightforward to build; Google's free recommendation systems course\footnote{https://developers.google.com/machine-learning/recommendation} provides an estimate that the course will take a mere four hours to complete (given prerequisite skills).  Because of the ease of implementing a basic system, algorithmic recommendations are prolific online, deployed by business large and small to impact millions of people each day.\looseness=-1

Recently, there has been increasing interest in understanding the impact of personalized recommendation systems from perspectives that firms have historically not considered, such as: how much do these systems contribute to polarization?  Do they homogenize users?  Are they fair to all users or do they disadvantage portions of users?  These questions are difficult to answer from observational data and it may be risky, or even unethical, for firms to explore these ideas with A/B tests.  
Simulations provide an alternative approach to understand the impacts of recommendation systems; this is an increasingly popular approach to evaluating these systems and broad simulation tools are being developed \citep{lucherini2021t}.  Additionally, some recommendation methods, such as reinforcement learning techniques, require more sophisticated approaches to offline evaluation prior to moving to A/B testing; simulations are similarly appropriate in this and other contexts.\looseness=-1

When constructing a recommendation system simulation, there are two key challenges. 
The first is defining a model of user choice.  There are a variety of well-established consumer choice models from economics and marketing, which will be described in more detail in \Cref{sec:consumer_choice_models}.  In accordance with the notion that ``all models are wrong,''  the true nature of consumer decision-making is likely more complicated and heterogeneous than defined by these choice models. 
That said, these models of choice offer us hope: if machine learning models can capture behavior patterns under these simplified assumptions, we have a chance of recovering and understanding users' true preferences in a more nuanced world. This would allow us to to begin describing the limitations of machine learning recommendation systems (e.g., by answering the previously posed questions about their impact) and therefore use them more judiciously in practice. 

The other key challenge is defining how users encounter items that are not recommended, including in the counterfactual world in which they do not receive algorithmic recommendations at all.  In the real world, consumers encounter content in a wide variety of ways (and sometimes even using a variety of similar platforms): search, promotions, lists of new content or content by theme, and recommendations from friends or experts are a few alternative mechanisms.
A simulation that does not include the notion of users accessing items in alternate ways will overestimate the effects of algorithmic recommendations on user behavior. This second challenge is not wholly disentangled from the first one of defining a model of user choice, but we will address it distinctly in \Cref{sec:alternarives_to_rec}.

While there are undoubtedly a variety of other challenges associated with building a recommendation system simulator, this position paper will delve into both of these key challenges, reviewing simulation assumptions for existing research in this context, as well as proposing alternative assumptions worth further exploration.  We will conclude with a broader discussion limitations of simulations and an outline of open questions in this area (\Cref{sec:discussion}).

\section{Models of Consumer Choice}
\label{sec:consumer_choice_models}

In order for users to select items from among the recommendations, a simulator must include some model of user choice; this is our first key challenge.  Existing work in recommendation system simulations typically represents both user preferences and items attributes in some relatively low-dimensional space, not unlike the very recommendation systems they seek to evaluate.  For example, the general purpose T-RECS tool \citep{lucherini2021t} uses vector representations for each item and user, with the ``score'' of an item for a user being either the inner product between the user preferences and item attributes, or the cosine similarity between the two representations.  Even within this construction, however, there are a wide variety of assumptions about how users make their choices.

Some simulations assume that users are uncertain of their preferences.  \citet{aridor2020deconstructing} model users as considering items sequentially, updating their beliefs as they go; under this paradigm, recommendations amount to adjusting the values which users rely on for decisions.
Other work frames the utility of an item for a user as being comprised of two parts: known and unknown to the user; a user then acts based on a function of the recommended rank of an item and their known utility for that item \citep{chaney2018algorithmic}.  Still other simulations forgo the individual element, assuming a universal quality value for each item with items being selected based on a function of eitehr rank (popularity) or quality based on a popularity bias parameter \citep{ciampaglia2018algorithmic}.
Sometimes users are represented as points in a low-dimensional attitude space with items being selected based on distance metrics \citep{geschke2019triple}.
\citet{yao2017beyond} use stochastic block models to generate data: users probabilistically belong to groups and users within a group probabilistically like items.
Unfortunately, some work does not make the exact choice model clear, specifying only that users select a subset of recommended items and provide feedback on this set
\citep{jiang2019degenerate}; the mechanism for selection is unspecified.  In other cases, simulations assume that users act on every single recommendation, rating each according to their preferences \citep{sun2019debiasing}.\looseness=-1

Regardless of the choice model used, researchers need to be explicit about these assumptions as it is important for replicability; simple alterations in assumptions may drastically alter results.  In addition, it would be beneficial to work towards a set of standard choice models.  In doing so, I encourage us to look to consumer choice models from economics and marketing.  Relying on these existing models confers multiple benefits to researchers and analysts: these models have been critiqued extensively, theoretically vetted, used to analyze real-world behaviors in multiple contexts, and bring with them established terminology; we need not reinvent the wheel.  While no choice model will perfectly capture all the nuances in user behavior, it stands to reason that we should draw on and integrate with existing consumer choice models in defining how simulated users behave when interacting with recommendation systems.  To aid in this effort, we will discuss a few consumer choice models, adapting them to the context of recommendation systems.\looseness=-1

A fundamental premise shared among consumer choice models is that consumers, or users, make choices based on the utility of different actions.  The utility $U_{nit}$ of user $n$ consuming an item $i$ (e.g., reading, watching, or purchasing it) at choice instance $t$ is typically comprised of two parts:  a deterministic function $f$ of the attributes of the item ($\mathbf{a}_i = \{a_{i1}, \dots a_{iJ}\}$ for $J$ attributes) and a random component $\epsilon$ specific to the choice instance $t$.
In computing a given overall utility $U_{nit}$, the utility of a given attribute can be unique for each user, giving us \looseness=-1
\begin{equation}
    \label{eq:general_utility}
    U_{nit} = f(\mathbf{a}_{i}, \mathbf{u}_{n}) + \epsilon_{nit},
\end{equation}
where $u_{nj}$ is the utility weight of attribute $j$ for user $n$.  The deterministic function $f$ is usually (but not always) linear: the deterministic utility of consuming an item is a linear combination of the utility weights of each of the item's attributes, or
\begin{equation}
    \label{eq:linear_utility}
    f(\mathbf{a}_{i}, \mathbf{u}_{n}) = \sum_{j=1}^J a_{ij} u_{nj}.
\end{equation}
This linear construction of utility has parallels with many of the simulation assumptions already used in recommendation system simulations, which draw from the matrix factorization model for recommendation \citep{koren2008factorization}.  It is also used in the seminal work of \citet{guadagni1983logit}, which introduced the multinomial logit model of choice; under this model, the random component $\epsilon$ is assumed to be drawn from a standard Gumbel distribution.
As an alternative to this multinomial logit model, the multinomial probit model assumes a different distribution of the random component of the utility $\epsilon$; specifically, it assumes $\epsilon_{nit}$ is drawn from a multivariate normal $\boldsymbol{\epsilon}_{nt} \sim \mathcal{N}(0, \boldsymbol{\Sigma})$, allowing for arbitrary correlation.  In both cases, consumers maximize their utility in expectation with the probability of choosing an item being a function of all the item alternatives available $\mathcal{I}_t$ at a given choice instance~$t$.  In the simulation setting, we use behavior models to \emph{generate} data rather than to describe observed behaviors, so to use this model in simulating choices we need only draw the collection of random components $\boldsymbol{\epsilon}$ from the appropriate distributions; each user $u$  then selects the item $i$ that maximizes their utility $U_{nit}$ for a given choice instance $t$.  However, care must be taken in choosing the magnitude of user utility weights $\mathbf{u}_n$ relative to the random components $\boldsymbol{\epsilon}$;  if the deterministic component of utility is too large, then there will be no randomness in consumer behavior, but if it is too small, user behavior will be exclusively random.

Personalized recommendation systems are only used in domains where users select multiple items, meaning the models of choice we consider must accommodate multiple selection.  The linear multinomial logit and probit model just described easily integrate this constraint but allow for the same item to be selected multiple times.  In some contexts, however, this may not be a desirable property.  For these cases, we may modify the models to exclude items previously selected by the user $n$ from consideration, giving us a user-specific set item alternatives $\mathcal{I}_{nt}$.  These two sets of assumptions represent extreme ends of a spectrum: repeated consumption of the same item could yield identical utility in expectation or zero utility after the first instance of consumption.  In many domains, the reality for consumers would be a hybrid of the two: \emph{diminishing} utility for repeat consumption.  For simplicity, one could consider only the two extremes and argue that a diminishing utility model would perform somewhere in between the two.\looseness=-1

When consumers can easily consider all items, the model assumptions discussed thus far seem appropriate.  Unfortunately, recommendation systems are deployed in exactly the context in which users \emph{cannot} consider all items.
Because of this context, the order in which users consider items matters, which brings us to the notion of ``satisficing'' choice~\citep{simon1955behavioral}.  This gives us a choice model wherein a consumer picks first item to meet a minimum threshold.
Under a the model proposed by \citet{stuttgen2012satisficing}, users have an ``aspiration level'' for each item attribute: this is the point at which the item would be considered satisfactory; for an item $i$ to be satisfactory overall, each attribute $j$ needs to be satisfactory.  With a large number of attributes, however, it is unrealistic for consumers to consider each attribute individually. Instead, we have user $n$ accept the first item $i$ with utility $U_{nit}$ (\Cref{eq:general_utility}) greater than or equal to a user-specific threshold, or overall aspiration level $\alpha_n$.  These aspiration levels may be fixed or learned by the user as a function of their experiences on the platform.  Anecdotally, this may be the best model for simulating consumer choice in the recommendation system context due to its simplicity.  However, like with determining the magnitude of utility weights $\mathbf{u}_n$, care must be taken in choosing the aspiration thresholds $\alpha_n$; too large thresholds yield minimal user interactions and overemphasize deterministic preferences whereas too small thresholds generate a large number of highly random user interactions.\looseness=-1

In addition to the utilities of the items themselves, we can incorporate the cost of thinking \citep{shugan1980cost} or search costs \citep{stigler1961economics} into our choice models.  This leads to another family of models based on consideration sets \citep{hauser1990evaluation, roberts1991development}. In the words of \citet{hauser1990evaluation}, these models capture the ``trade-offs between decision costs and the incremental benefits of choosing from a larger set.''  Given a consideration set of items $\mathcal{C}$, a consumer $n$ must choose between evaluating the items in the consideration set to select one, or searching for a new item to consider, with search cost $s_n$.  If they choose to search and discover a new item, they must determine if it is worth the decision cost $d_n$ to add the item to the consideration set.  Items may also be dropped from a consideration set if their contributions to the expected utility of the final choice do not outweigh the decision cost $d_n$ of including them in the set.  One could assume that each user $n$ has their own decision costs $d_n$ and search costs $s_n$ that are constant.  Under this model, each user $n$ starts with an empty consideration set $\mathcal{C}_n$ and selects items using the following procedure, based on \citet{hauser1990evaluation} for a given choice instance $t$.\looseness=-1
\vspace{-2.5px}
\begin{itemize}
    \item Drop any item $i$ from the consideration set $\mathcal{C}_n$ if its expected contribution to utility is not worth the decision cost, or $ \mathbb{E}_t \left[\max_{i' \in \mathcal{C}_n} U_{ni't}\right] - \mathbb{E}_t \left[\max_{i' \in \mathcal{C}_n \setminus \left\{i\right\}} U_{ni't}\right] < d_n$.
    \item 
    If the expected marginal utility of considering a new item with the associated decision costs is less than the search cost, or $\mathbb{E}_{t,i}\left[\max_{i' \in \mathcal{C}_n \cup \left\{i\right\}}  U_{ni't}\right] - \mathbb{E}_t\left[\max_{i' \in \mathcal{C}_n}  U_{ni't}\right] + d_n < s_n$, search for a new item $i \in \mathcal{I}_{nt}$ and continue.
    Otherwise, return to the start of the procedure after choosing the highest utility item from $\mathcal{C}_n$, $y_{nt} = \arg \max_{i \in \mathcal{C}_n} U_{nit}$, and then removing $y_{nt}$ from $\mathcal{C}_n$.
    \item Add item $i$ to the consideration set $\mathcal{C}_n$ if the expected marginal utility of adding the item is greater than the decision cost,
    $\mathbb{E}_{t}\left[\max_{i' \in \mathcal{C}_n \cup \left\{i\right\}}  U_{ni't}\right] - \mathbb{E}_t\left[\max_{i' \in \mathcal{C}_n}  U_{ni't}\right] > d_n$.
\end{itemize}
\vspace{-2.5px}
In practice, this model requires extensive computation to estimate all the user-specific expected contributions and marginal utilities.  Once these challenges are overcome, this could be a potentially powerful model for simulating user choices.

In all of the consumer choice models discussed here, we have assumed that deterministic consumer preferences are static except for small random variations $\epsilon$; these deterministic preferences do not evolve systematically in any way.  The satisfying model includes some notion of adapting preference if we allow the aspirational thresholds $\boldsymbol{\alpha}$ to be learnt; the search model similarly includes evolving preferences based on expected marginal utilities, which are dynamic based on the users' experiences.  However, there is more room to explore the dynamics of evolving consumer preferences in the recommendation systems context; for example, it is established that consumers can learn their own preference weights $\mathbf{u}_n$ as they search, complicating the process of recommendation \citep{dzyabura2019recommending}.  These and similar complex dynamics have been explored in economics and marketing, providing recommendation system researchers a plethora of models from which to draw inspiration for their simulation assumptions regarding user choices.

\section{Alternatives to Recommendations}
\label{sec:alternarives_to_rec}

We now turn to an even broader challenge: in the real world, users interact with items through a combination of online and offline mechanisms, including, but not limited to, algorithm recommendation.  For example, a consumer may receive a book recommendation from a friend in conversation and then use a search engine to find it later.  Users will always have alternatives to using a recommendation system, including leaving the platform entirely if it does not meet their needs.
Excluding this general concept from the simulation setting will vastly overemphasize the impact of any given recommendation system on user behavior.
To date, recommendation system simulations address this in simple ways, if at all.   \looseness=-1 

Some work offers no alternatives to recommendations, focusing only on generating data with bias and the corresponding implications \citep{yao2017beyond,sun2019debiasing}.
Others compare different recommendation systems such as random recommendations, recommendations based on overall popularity, matrix factorization, and recommendations under ideal or oracle conditions \citep{chaney2018algorithmic, jiang2019degenerate, aridor2020deconstructing}. Along this vein, \citet{geschke2019triple} compares recommending close content to distant content, but also includes social dynamics in their simulations.  Even more simply, some simulations interleave random items among recommended ones \citep{chaney2018algorithmic}.\looseness=-1

Many researchers dismiss the need for alternatives to recommended content or re-frame their questions to avoid this difficult issue.  While this is an understandable starting point, we need to move past it.   An observational study of Bing search and Amazon data estimated that at least 75\% of Amazon product browsing activity would likely occur in the absence of recommendations \citep{sharma2015estimating}.  While the exact amount of activity absent recommendations depends on the content and platform features, this example underscores the importance of alternate ways of accessing content.\looseness=-1

This second challenge is connected to the first one of defining a model of user choice: users consider not only which items to select, but the sources from which they select them.  On some level, then, there are easy solutions.  For example, a satisficing model presents users with an ordered list of items; in this context, we can extend the concept of interleaving random items to probabilistically selecting items for consideration from different sources.  Items may be algorithmically recommended, but they may also be new items, items returned by search, etc.  Ideally, these probabilities would change as a function of the learned relevancy for each source, not unlike a multi-armed bandit model.  The consideration set model previously discussed can similarly draw on multiple sources with distinct search costs for each source.
The true difficulty is that each of these alternate mechanisms, or sources, needs a corresponding model with its own assumptions.\looseness=-1

Again, we need standards for alternate mechanisms for accessing content in simulations, including the associated assumptions with each mechanism.  This may involve observational research to understand the breadth of how users access content and to characterize the nature of their interactions with these alternate sources.

\section{Discussion}
\label{sec:discussion}

As simulation methods become more popular for understanding the impact of recommendation systems, we must be critical of the assumptions we use and work toward establishing standards based on real-world consumer behaviors.  This includes clear definitions of our user choice models (\Cref{sec:alternarives_to_rec}) and realistic mechanisms for accessing content as alternatives to algorithmic recommendations (\Cref{sec:alternarives_to_rec}).  In considering the impacts of recommendation systems, we need to define what the comparison is, or what can realistically be changed in the world.  
Firms cannot deploy an oracle recommendation system, though they undoubtedly wish they could.  Instead, the data that they have is biased by their existing recommendation system or platform design choices; they need to make the best choices they can within those constraints.  Our work should focus on realistic comparisons that lead to actionable results for firms.  This includes how to handle new items and users (the ``cold-start'' problem), understating the benefits and consequences of bias reduction strategies, how to best combine historic and recent data, and comparing pipeline choices such as offline evaluation and A/B testing metrics, as well as the frequency of retraining or updating the system.  
We also need standards on the scale of simulations; real-world platforms are large with thousands if not millions of users.  What size of simulations are needed to generate compelling results?  How does the required scale vary based on the simulation assumptions?\looseness=-1

While these challenges may seem daunting, they are part of a larger, enduring task.  As our society continues to integrate with technical systems like recommendation engines, we must do our best to understand the impacts of these systems through simulations and other methods so we may use these technologies wisely.\looseness=-1

\begin{acks}
Thank you to Brandon Stewart, Ryan Dew, Sam Levy, the marketing faculty at both Chicago Booth and Duke Fuqua, and my fellow organizers of the SimuRec workshop for their discussions broadly related to this work.
\end{acks}

\bibliographystyle{ACM-Reference-Format}
\bibliography{bibliography}

\end{document}